\documentclass[10pt]{iopart}

\eqnobysec

\usepackage{iopams}
\usepackage{amstext}
\usepackage{placeins}
\usepackage{setstack}
\usepackage{graphicx}

\begin{document}

\title[Fine structure of absorption cross sections for black holes]
{Fine structure of high-energy absorption cross sections for black holes}

\author{Yves D\'ecanini}
\address{Equipe Physique Th\'eorique, SPE, UMR 6134 du CNRS
et de l'Universit\'e de Corse,  Universit\'e de Corse, Facult\'e des
Sciences, BP 52, F-20250 Corte, France}
\author{Antoine Folacci}
\address{Equipe Physique
Th\'eorique, SPE, UMR 6134 du CNRS et de l'Universit\'e de Corse,
Universit\'e de Corse, Facult\'e des Sciences, BP 52, F-20250 Corte,
France \\ and \\Centre de Physique Th\'eorique, UMR 6207 du CNRS et
des Universit\'es Aix-Marseille 1 et 2 et de l'Universit\'e du Sud
Toulon-Var, CNRS-Luminy Case 907, F-13288 Marseille, France}
\author{Bernard Raffaelli}
\address{Equipe Physique
Th\'eorique, SPE, UMR 6134 du CNRS et de l'Universit\'e de Corse,
Universit\'e de Corse, Facult\'e des Sciences, BP 52, F-20250 Corte,
France} \eads{\mailto{decanini@univ-corse.fr} and
\mailto{folacci@univ-corse.fr} and \mailto{raffaelli@univ-corse.fr}}

\begin{abstract}

\noindent The high-energy absorption cross section of the
Schwarzschild black hole is well approximated, in the eikonal
regime, by the sum of two terms: the geometrical cross section of
the black hole photon sphere and the contribution of a sinc function
involving the geometrical characteristics (orbital period and
Lyapunov exponent) of the null unstable geodesics lying on this
photon sphere. From a numerical analysis, we show that, beyond the
eikonal description, this absorption cross section presents a simple
fine structure. We then describe it analytically by using Regge pole
techniques and interpret it in geometrical terms. We naturally
extend our analysis to arbitrary static spherically symmetric black
holes endowed with a photon sphere and we then apply our formalism
to Schwarzschild-Tangherlini and Reissner-Nordstr\"om black holes.
Finally, on the example of the Schwarzschild black hole, we show
numerically that a complicated hyperfine structure lying beyond the
fine structure can also be observed.

\end{abstract}


% \submitto{\CQG}

\section{Introduction}

In a recent paper \cite{DecaniniEspositoFareseFolacci2011}, by using
Regge pole techniques, we have developed a new and universal
description of the absorption problem for a massless scalar field
propagating in static spherically symmetric black holes of arbitrary
dimension endowed with a photon sphere. We have shown, in
particular, that the high-energy absorption cross section is well
approximated, in the eikonal regime, by the sum of two
contributions: the geometrical cross section of the black hole
photon sphere (i.e., the so-called capture cross section of the
black hole) and a sinc function involving the geometrical
characteristics (orbital period and Lyapunov exponent) of the null
unstable geodesics lying on this photon sphere. We have therefore
provided a rigorous analysis as well as a clear physical description
of a result well known in black hole physics (see, e.g.,
Refs.~\cite{Sanchez1978a,
HarrisKanti2003,JungPark2004,JungPark2005,DoranEtAL2005,
GrainBarrauKanti2005,CrispinoEtAL2007,DolanOliveiraCrispino2009,
CrispinoDolanOliveira2009}), the fact that in general, at high
energies, the absorption cross section of a black hole oscillates
around a limiting constant value, explaining within the same
formalism the existence of this limiting value and of the
fluctuations.

\bigskip

In this paper, from the complex angular momentum formalism already
developed in Ref.~\cite{DecaniniEspositoFareseFolacci2011} and by
using new asymptotic expansions for the residues of the ``greybody"
factors, we shall go beyond the eikonal description of the
high-energy absorption cross section for black holes. More
precisely, in section 2, from a numerical analysis, we shall first
show that, beyond the eikonal description, the absorption cross
section of the Schwarzschild black hole presents a simple fine
structure. We shall then describe it analytically and interpret it
in geometrical terms. In section 3, we shall extend the previous
considerations to static spherically symmetric black holes endowed
with a photon sphere and apply the formalism developed to the five-
and six-dimensional Schwarzschild-Tangherlini black holes and to the
four-dimensional Reissner-Nordstr\"om black hole. In a brief
conclusion (our section 4), we shall consider some possible
consequences of our present work and show numerically, for the
Schwarzschild black hole, the existence of a complicated hyperfine
structure lying beyond the fine structure. Finally, in a much more
technical appendix, we shall provide some asymptotic expansions for
the Regge poles and for the corresponding residues of the greybody
factors which are useful to describe analytically the fine structure
and which could be helpful to analyze the hyperfine structure.

\bigskip

Throughout this paper, we shall use units such that $\hbar = c = G =
1$ and we shall assume a harmonic time dependence $\exp(-i\omega t)$
for the fields.

\section{Fine structure of the high-energy absorption cross sections of the
Schwarzschild black hole}

We begin by various considerations relative to (i) the Schwarzschild
black hole and (ii) a scalar field theory defined on this
gravitational background that we shall use extensively in this
section and in subsection A.1 of the appendix. This moreover permits
us to establish all our notations:

\smallskip
\qquad (i) We first recall that the exterior of the Schwarzschild
black hole of mass $M$ is usually defined as the manifold ${\cal M}
= \hbox{$]-\infty, +\infty[_t$} \times ]2M,+\infty[_r \times S^2$
with metric $ds^2= -f(r)dt^2+ f(r)^{-1}dr^2+ r^2d\sigma_2^2.$ Here
$d\sigma_2^2$ denotes the metric on the unit $2$-sphere $S^2$ and
$f(r)=(1-2M/r)$. We remark that, instead of the standard radial
Schwarzschild coordinate $r$, it is sometimes more convenient to use
the so-called tortoise coordinate $r_\ast(r)$ defined by
$dr/dr_\ast=f(r)$ which provides a bijection from $]2M,+\infty[$ to
$]-\infty,+\infty[$. We also note that this black hole has a photon
sphere located at $r=3M\equiv r_c$, that the corresponding critical
impact parameter is given by $b_c= 3\sqrt{3} M$ (see, e.g., Chap.~25
of Ref.~\cite{MTW}) and that, as a consequence, the geometrical
cross section of this black hole is $\sigma_\mathrm{geo}=\pi
b_c^2=27\pi M^2$.

\smallskip
\qquad (ii) We also recall that the wave equation for a massless
scalar field propagating on the Schwarzschild black hole reduces,
after separation of variables and the introduction of the radial
partial wave functions $\phi_{\omega \ell}(r)$, to the Regge-Wheeler
equation
\begin{equation}\label{RW}
\frac{d^2 \phi_{\omega \ell}}{dr_*^2} + \left[ \omega^2 -
V_\ell(r_\ast)\right] \phi_{\omega \ell}=0
\end{equation}
where the so-called Regge-Wheeler potential $V_\ell(r_\ast)$ is
given in terms of the radial Schwarzschild coordinate by
\begin{equation}\label{EffectivePot_spin_0}
V_\ell(r) = \left(\frac{r-2M}{r} \right) \left[
\frac{(\ell+1/2)^2-1/4}{r^2} +\frac{2M}{r^3}\right].
\end{equation}
Here $\omega>0$ denotes the frequency of the mode-solution
considered while $\ell \in \mathbb{N}$ is the ordinary angular
momentum index. We note that for $\omega$ and $\ell$ given, the
radial amplitude $\phi_{\omega \ell}(r_\ast)$ moreover satisfies the
boundary conditions
\begin{equation}\label{BC_Schw}
\phi_{\omega \ell}(r_\ast) \sim \left\{
\begin{array}{cl} T_\ell(\omega) e^{-i\omega r_\ast}&
\mathrm{for} \ \
r_* \to -\infty,\\
&\\
e^{-i\omega r_\ast} + R_\ell(\omega) e^{+i\omega r_\ast}&
\mathrm{for} \ \ r_* \to +\infty,
\end{array}
\right.
\end{equation}
where $T_\ell(\omega)$ and $R_\ell(\omega)$ are transmission and
reflection coefficients. We finally recall that the greybody factors
(the absorption probabilities for particles with energy $\omega$ and
angular momentum $\ell$) are given by
\begin{equation}\label{Greybodyfactors}
\Gamma_\ell(\omega)= |T_\ell(\omega)|^2
\end{equation}
and that, for the massless scalar field considered here, the black
hole absorption cross section can be expressed in terms of the
greybody factors in the form
\begin{equation}\label{Sigma_abs}
\sigma_\mathrm{abs}(\omega)=\frac{\pi}{\omega^2}
\sum_{\ell=0}^{+\infty} (2\ell + 1) \Gamma_\ell(\omega).
\end{equation}
It should be noted that the series (\ref{Sigma_abs}) can be
evaluated with a very great precision by solving numerically the
problem defined by (\ref{RW}), (\ref{EffectivePot_spin_0}) and
(\ref{BC_Schw}) but that its evaluation is very time consuming for
high frequencies.

\bigskip

\begin{figure}
\includegraphics[width=13.1cm]{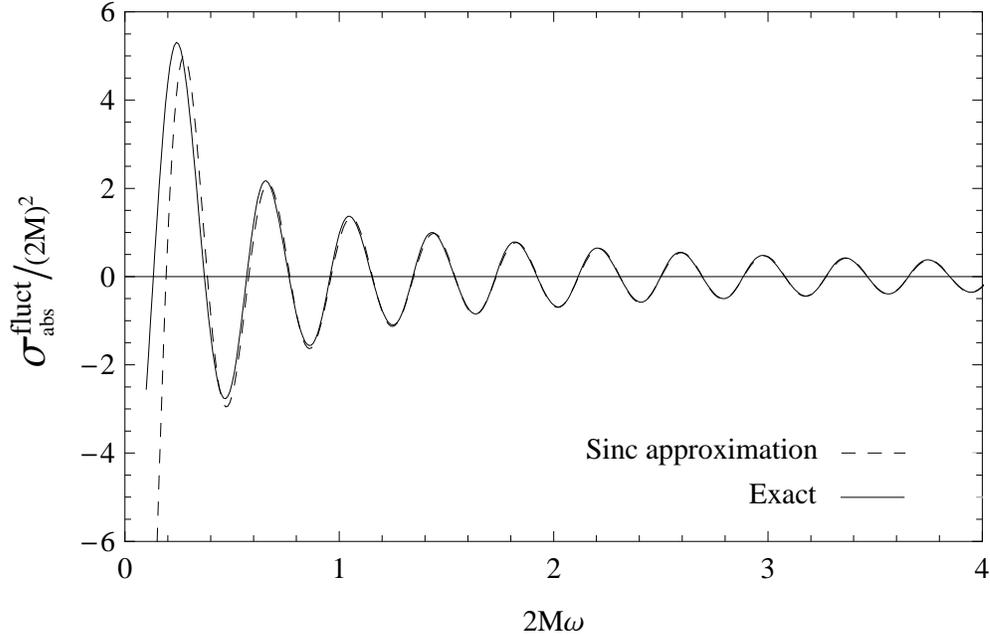}
\noindent \caption{\label{fig:SecAbsExactAndEikonale}Fluctuations of
the total absorption cross section,
$\sigma^\mathrm{fluct}_\mathrm{abs} \equiv \sigma_\mathrm{abs} -
\sigma_\mathrm{geo}$, for a massless scalar field propagating in the
Schwarzschild geometry. The exact curve is obtained numerically from
(\ref{Sigma_abs})  while the sinc approximation is given by
(\ref{Sigma_abs_PR_a_APP}).}
\end{figure}

\begin{figure}
\includegraphics[width=13.1cm]{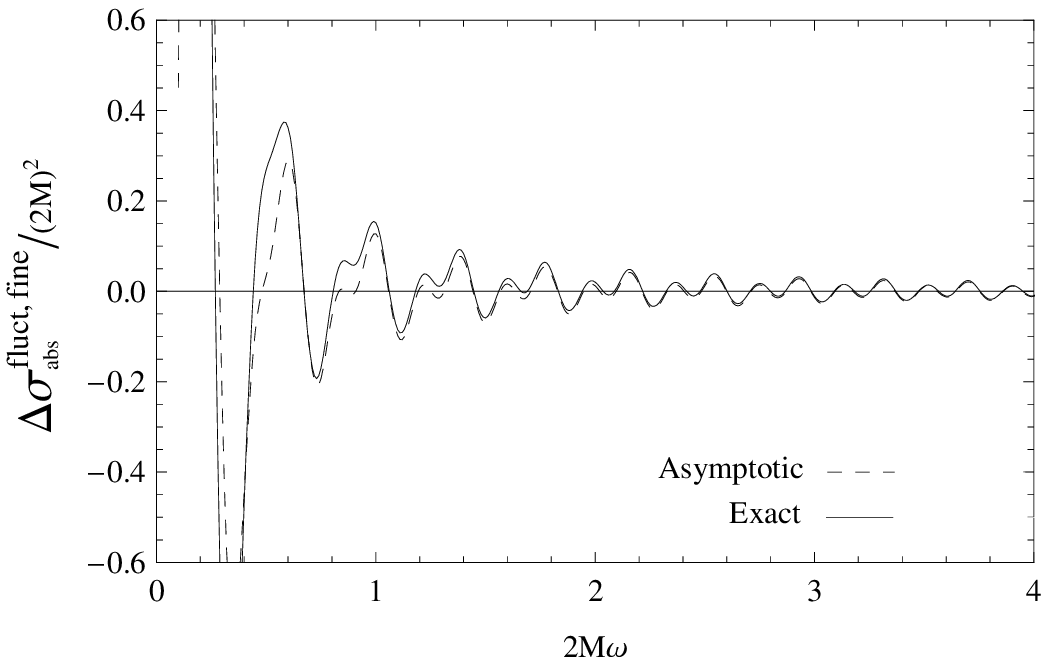}
\caption{\label{fig:FineStructure}Fine structure of the total
absorption cross section, $\Delta \sigma_\mathrm{abs}^\mathrm{fluct,
\, fine}(\omega) \equiv \sigma_\mathrm{abs}(\omega) -\left[
\sigma_\mathrm{geo}+\sigma_\mathrm{abs}^\mathrm{osc}(\omega)
\right]$, for a massless scalar field propagating in the
Schwarzschild geometry. The exact curve is obtained numerically from
(\ref{Sigma_abs}) and (\ref{Sigma_abs_PR_a_APP}) while the
asymptotic result is given by (\ref{Struct_fine_asymp}).}
\end{figure}

In Ref.~\cite{DecaniniEspositoFareseFolacci2011}, by using Regge
pole techniques, we have proved that the high-energy behavior of the
absorption cross section of the Schwarzschild black hole
(\ref{Sigma_abs}) is well approximated, in the eikonal regime, by
the sum of two terms (what we shall call from now on the eikonal
description): the geometrical cross section of the black hole photon
sphere and a sinc function involving the geometrical characteristics
(orbital period and Lyapunov exponent) of the null unstable
geodesics lying on this photon sphere (see also
Ref.~\cite{Sanchez1978a} for a fit of the absorption cross section
of the Schwarzschild black hole involving a sinc function but based
mainly on numerical considerations). More precisely, we have
rigorously shown that the fluctuations of (\ref{Sigma_abs}) around
the geometrical cross section $\sigma_\mathrm{geo}$ are well
described by the very simple formula
\begin{equation}\label{Sigma_abs_PR_a_APP}
\sigma_\mathrm{abs}^\mathrm{osc}(\omega)=-8 \pi\, e^{-\pi}\,
\sigma_\mathrm{geo} \, \mathrm{sinc}\left[2\pi (3\sqrt{3}
M)\omega\right],
\end{equation}
where $\mathrm{sinc}\, x \equiv (\sin x)/x$ is the sine cardinal.
Let us note that the argument of the sinc involves the orbital
period, $2\pi (3\sqrt{3} M)=2\pi b_c$, of a massless particle
orbiting the black hole on the photon sphere
\cite{DecaniniFJ_cam_bh,DecaniniFolacci2010a,DecaniniFolacciRaffaelli2010b,
CardosoMirandaBertietal2009} while the coefficient $8 \pi e^{-\pi}$
is linked to the Lyapunov exponent of the geodesic followed by the
particle (see
Refs.~\cite{DecaniniEspositoFareseFolacci2011,DecaniniFolacciRaffaelli2010b,
CardosoMirandaBertietal2009} for more precisions on this last
point). This formula permits us to interpret naturally the period of
the maxima (or the minima, or the zeros) of the fluctuations in
terms of constructive interferences of the ``surface waves" trapped
near the photon sphere (see also Refs.~\cite{DecaniniFJ_cam_bh} and
\cite{DecaniniFolacci2010a} for related aspects).

\bigskip

The agreement of (\ref{Sigma_abs_PR_a_APP}) with the exact result
obtained numerically from (\ref{Sigma_abs}) is very good, even for
low frequencies (see figure \ref{fig:SecAbsExactAndEikonale}).
However, if we consider the difference between the two curves
displayed, i.e., if we consider the behavior of the function
\begin{equation}\label{Struct_fine_exact}
\Delta \sigma_\mathrm{abs}^\mathrm{fluct, \, fine}(\omega) \equiv
\sigma_\mathrm{abs}(\omega) -\left[
\sigma_\mathrm{geo}+\sigma_\mathrm{abs}^\mathrm{osc}(\omega)
\right],
\end{equation}
we show that beyond the eikonal description, the absorption cross
section presents a simple fine structure (see figure
\ref{fig:FineStructure}). Because the amplitude of this fine
structure is around $5 - 10 \, \%$ of that of the eikonal
fluctuations, its existence must be mentioned and could even have
interesting physical consequences. So it seems to us worthwhile to
describe it mathematically.

\bigskip

With this aim in mind, let us first recall how we obtained the
eikonal description of the absorption cross section
(\ref{Sigma_abs}). In Ref.~\cite{DecaniniEspositoFareseFolacci2011},
from the Regge pole machinery, we have shown that (\ref{Sigma_abs})
can be replaced by the series
\begin{eqnarray}\label{sigma_abs_SC3}
\fl \sigma_\mathrm{abs}(\omega)=27\pi M^2
-\frac{4\pi^2}{\omega^2}\,\mathrm{Re} \left( \sum_{n=1}^{+\infty}
\frac{ e^{i\pi[\lambda_n(\omega)-1/2]}\, \lambda_n(\omega)
\gamma_n(\omega)} {\sin[\pi (\lambda_n(\omega)-1/2)]} \right) +
\underset{M\omega \to +\infty}{\cal
O}\left(\frac{1}{(M\omega)^2}\right). \nonumber \\
\fl
\end{eqnarray}
In equation (\ref{sigma_abs_SC3}), the $\lambda_n(\omega)$ with $n
\in \mathbb{N}\setminus\lbrace{0\rbrace}$ are those of the (Regge)
poles of the analytic extension $\Gamma_{\lambda-1/2}(\omega)$ of
the greybody factor $\Gamma_\ell(\omega)$ lying in the first
quadrant of the complex $\lambda$ plane and the $\gamma_n(\omega)$
are the associated residues. It is important to recall that
(\ref{sigma_abs_SC3}) converges very rapidly and that the
contribution of the Regge poles with $n >1$ is practically
negligible (see figure 2 of
Ref.~\cite{DecaniniEspositoFareseFolacci2011}). So, by using the
rough approximations
\begin{equation}\label{PRapproxWKB_rough}
\lambda_n(\omega) = 3\sqrt{3}M\omega + i \left(n-1/2 \right) +
\underset{M\omega \to +\infty}{\cal O}\left(\frac{1}{M\omega}\right)
\end{equation}
and
\begin{equation}\label{RESapproxWKB_rough}
\gamma_n(\omega) = -\frac{1}{2\pi} + \underset{M\omega \to
+\infty}{\cal O}\left(\frac{1}{M\omega}\right),
\end{equation}
and taking into account only the contribution of the first Regge
pole, we obtained from (\ref{sigma_abs_SC3}) the eikonal
approximation (\ref{Sigma_abs_PR_a_APP}) for the fluctuations of the
absorption cross section (\ref{Sigma_abs}).

\bigskip

Of course, approximations (\ref{PRapproxWKB_rough}) and
(\ref{RESapproxWKB_rough}) are too rough to permit us to understand
the existence of the fine structure. But if we now consider the
asymptotic expansions (\ref{PRapproxWKB_hf3}) and
(\ref{RESapproxWKB_hf3}) with $n=1$ and $s=0$, by using the relation
(for $a \in {\mathbb {R}}$)
\begin{equation}
\frac{e^{i \pi (z-a)}}{\sin [\pi (z-a)] } =-2i\sum_{m=1}^{+\infty}
e^{i 2m \pi (z -a)}\quad  \mathrm{valid \,\, if}\, \, \mathrm{Im} \
z > 0, \label{devS sin_a}
\end{equation}
we can show from (\ref{sigma_abs_SC3}) that (\ref{Sigma_abs}) can be
approximated by
\begin{eqnarray} \label{Struct_fine_asymp}
\fl \sigma_\mathrm{abs}(\omega) \approx \sigma_\mathrm{geo}\left(1
-8 \pi\, e^{-\pi}\,  \frac{\sin \left[2\pi (3\sqrt{3}
M)\omega\right]}{2\pi (3\sqrt{3}
M)\omega}  \right.\nonumber \\
\fl \left.\qquad + 16\pi e^{-2\pi}
\frac{\sin{[4\pi(3\sqrt{3}M)\omega]}}{4\pi(3\sqrt{3}M)\omega}
+\frac{4\pi^2 e^{-\pi}(-39+7\pi)}{27}
\frac{\cos{[2\pi(3\sqrt{3}M)\omega]}}{[2\pi(3\sqrt{3}M)\omega]^2}\right).
\end{eqnarray}
In (\ref{Struct_fine_asymp}), the first two terms correspond to the
eikonal description constructed in
Ref.~\cite{DecaniniEspositoFareseFolacci2011} while the third and
fourth ones describe the fine structure of the absorption cross
section. In figure \ref{fig:FineStructure} we have compared the
exact fine structure numerically evaluated with the result provided
by the third and fourth terms of (\ref{Struct_fine_asymp}). The
agreement is truly remarkable. In fact, the error made on the exact
absorption cross section (\ref{Sigma_abs}) or on its fluctuations
around $\sigma_\mathrm{geo}$ is considerably reduced by using
(\ref{Struct_fine_asymp}) (see also figure
\ref{fig:HyperFineStructure} and the associated comment in section
4).

\bigskip

It is now important, from a physical point of view, to note that the
fine structure, as the eikonal contribution, is only due to the
``surface wave" trapped near the photon sphere which is associated
with the first Regge pole (for the interpretation of Regge poles in
terms of ``surface waves", we refer to
Refs.~\cite{Andersson2,DecaniniFJ_cam_bh,DecaniniFolacci2010a,
DecaniniFolacciRaffaelli2010b}. However, in order to describe the
fine structure, we must now take into account (i) the multiple
circumnavigations around the black hole of this surface wave (the
``beats" in the fine structure observed in figure
\ref{fig:FineStructure} are due to interferences between terms
involving the orbital period of a massless particle orbiting the
black hole on the photon sphere and its second harmonic) and (ii)
its dispersive character (the amplitude of the first harmonic
contribution is constructed, in part, from the nonlinearities of the
first Regge pole trajectory, i.e., comes from the term in
$1/(M\omega)$ of (\ref{PRapproxWKB_hf3})).

\bigskip

To conclude this section, we would like to remark that with equation
(\ref{Struct_fine_asymp}) we have at our disposal a rather simple
and very accurate approximation permitting us to describe
qualitatively and quantitatively the high-energy behavior of the
absorption cross section of the Schwarzschild black hole and to thus
avoid very time-consuming calculations. It is moreover interesting
to note that, for ``very high energies", only the first three terms
of (\ref{Struct_fine_asymp}) must be taken into account what
simplifies considerably this approximation due to the elimination of
a rather inelegant term.

\section{Fine structure of high-energy absorption cross sections
for static spherically symmetric black holes}

\subsection{General theory}

The analysis developed in the previous section can be naturally
extended to the more general case of a massless scalar field theory
defined on a static spherically symmetric black hole of arbitrary
dimension $d \ge 4$ endowed with a photon sphere. The exterior of
such a black hole can be defined as the manifold ${\cal M} =
\hbox{$]-\infty, +\infty[_t$} \times ]r_h,+\infty[_r \times S^{d-2}$
with metric $ds^2= -f(r)dt^2+ f(r)^{-1}dr^2+ r^2d\sigma_{d-2}^2$
where $d\sigma_{d-2}^2$ denotes the metric on the unit
$(d-2)$-sphere $S^{d-2}$. Here the (standard) coordinate $r_h$ of
the event horizon is assumed to be a simple root of $f(r)$. We
furthermore assume that we have $f(r)>0$ for $r > r_h$ and
$\underset{r \to +\infty}{\lim}f(r)=1$. In other words, the
gravitational background considered is an asymptotically flat one
and the tortoise coordinate $r_\ast(r)$ defined again by
$dr/dr_\ast=f(r)$ provides a bijection from $]r_h,+\infty[$ to
$]-\infty,+\infty[$. The existence of a photon sphere located at
$r_c \in \hbox{$]r_h,+\infty[$}$ is ensured if we finally assume
that the conditions $f'(r_c) - (2/r_c) f(r_c)=0$ and
$f''(r_c)-(2/r_c^2) f(r_c) <0$ are satisfied (see
Ref.~\cite{DecaniniFolacciRaffaelli2010b} for more details on these
last assumptions). It should be also noted that, for such a black
hole, the critical impact parameter and the corresponding
geometrical cross section are now given respectively by $b_c=
r_c/\sqrt{f(r_c)}$ and $\sigma_\mathrm{geo}=\pi^{(d-2)/2}
b_c^{d-2}/\Gamma(d/2)$ (see, e.g., Ref.~\cite{HarrisKanti2003}). Of
course, when $f(r)=1-2M/r$ and $d=4$ we recover all the results
concerning the Schwarzschild black hole we have listed at the
beginning of section 2.

\bigskip

In order to simplify various expressions appearing in this section
and in subsection A.2 of the appendix, we introduce the following
notations:
\begin{equation} \label{fc}
f_c\equiv f(r_c) \quad \mathrm{and} \quad f_c^{(p)} \equiv
f^{(p)}(r_c) \quad \mathrm{for} \quad p\ge 1
\end{equation}
and
\begin{equation}\label{Eta}
\eta_c \equiv \frac{1}{2}\sqrt{4f_{c}-2r_{c}^{2}f_c^{(2)}}.
\end{equation}
As already noted and discussed in
Ref.~\cite{DecaniniFolacciRaffaelli2010b}, the $\eta_c$ parameter
represents a kind of measure of the instability of the circular
orbits lying on the photon sphere and can be expressed in terms of
the Lyapunov exponent corresponding to these orbits introduced in
Ref.~\cite{CardosoMirandaBertietal2009}.

\bigskip

The wave equation for a massless scalar field propagating on this
gravitational background still reduces, after separation of
variables and the introduction of the radial partial wave functions
$\phi_{\omega \ell}(r)$ with $\omega >0$ and $\ell \in \mathbb{N}$,
to the Regge-Wheeler equation (\ref{RW}) but, instead of
(\ref{EffectivePot_spin_0}), we have
\begin{eqnarray}\label{RWPot_dimD}
&&V_\ell(r)=f(r) \left[
\frac{[\ell+(d-3)/2]^2-[(d-3)/2]^2}{r^2}\right.
\nonumber \\
&&\qquad \qquad \left.+ \frac{(d-2)(d-4)}{4r^2}f(r) +
\left(\frac{d-2}{2r}\right)f'(r) \right].
\end{eqnarray}
The boundary conditions (\ref{BC_Schw}) remain valid and the
greybody factors are still defined by (\ref{Greybodyfactors}) but,
now, the black hole absorption cross section is given by (see
Ref.~\cite{Gubser1997})
\begin{equation}\label{Sigma_abs_dimD}
\sigma_\mathrm{abs}(\omega)=\frac{\pi^{(d-2)/2}}{\Gamma
\left[(d-2)/2\right] \omega^{d-2}}  \sum_{\ell=0}^{+\infty}
\frac{(\ell+d-4)!}{\ell!} \left(2\ell + d-3\right)
\Gamma_\ell(\omega).
\end{equation}

\bigskip

In Ref.~\cite{DecaniniEspositoFareseFolacci2011} we have shown that
the fluctuations of (\ref{Sigma_abs_dimD}) around the geometrical
cross section are described by the Regge pole series
\begin{eqnarray}\label{Sigma_abs_PR_a_dimD}
&&\sigma_\mathrm{abs}^\mathrm{RP}(\omega)=
-\frac{4\pi^{d/2}}{\Gamma\left[(d-2)/2\right] \omega^{d-2}}\,
\mathrm{Re}
\left( \sum_{n=1}^{+\infty} \right. \nonumber\\
&&\left. \qquad  \frac{\Gamma[\lambda_n(\omega)+(d-3)/2]}
{\Gamma[\lambda_n(\omega)-(d-5)/2]}
\frac{e^{i\pi[\lambda_n(\omega)-(d-3)/2]}
\lambda_n(\omega)\gamma_n(\omega) }{\sin[\pi
(\lambda_n(\omega)-(d-3)/2)]} \right).
\end{eqnarray}
Of course, at first sight such a series, even if it provides an
exact description of the fluctuations, does not seem really
interesting from a physical point of view. In
Ref.~\cite{DecaniniEspositoFareseFolacci2011}, we have been able to
extract from it an eikonal approximation of the absorption cross
section based on rough approximations for the Regge poles and their
residues. In subsection A.2 of the appendix, we have obtained one
more order for these approximations so we can now go beyond the
eikonal description and construct the fine structure of the
absorption cross section. By using the asymptotic expansions
(\ref{RPHF}) and (\ref{Residu_SSSBH}) for the Regge poles
$\lambda_n(\omega)$ and the residues $\gamma_n(\omega)$, as well as
(\ref{devS sin_a}) and \begin{eqnarray} \label{dev_Gamma_sur_Gamma}
&&\frac{\Gamma(z+a)}{\Gamma(z+b)} \sim \left( \frac{1}{z}
\right)^{-a+b} \quad \mathrm{valid\,\, if}\,\, |z| \to +\infty \,\,
\mathrm{and} \,\, |\arg z| < \pi,
\end{eqnarray}
and keeping only the contribution of the first Regge pole in
(\ref{Sigma_abs_PR_a_dimD}), we obtain
\begin{eqnarray} \label{Struct_fine_asymp_dimD}
\fl \sigma_\mathrm{abs}(\omega) \approx
\sigma_\mathrm{geo}\left(1+(-1)^{d-3}\, 4(d-2) \pi\,\eta_c\, e^{-\pi
\eta_c} \frac{\sin\left[2\pi (r_c/\sqrt{f_c})\omega\right]}{2\pi
(r_c/\sqrt{f_c})\omega}\right.\nonumber \\
\fl \qquad \left.+8(d-2)\pi\eta_c e^{-2\pi\eta_c}
 \frac{\sin{[4\pi(r_c/\sqrt{f_c})\omega]}}{4\pi(r_c/\sqrt{f_c})\omega}\right.\nonumber \\
\fl \qquad \left.-(-1)^{d-3}4(d-2)\pi^2
e^{-\pi\eta_c}\left[a_c-(d-3)\eta_c^2-2\pi\eta_c a_1\right]
 \frac{\cos{[2\pi(r_c/\sqrt{f_c})\omega]}}{[2\pi(r_c/\sqrt{f_c})\omega]^2}\right.\nonumber \\
\fl \qquad \left.-16(d-2)\pi^2
e^{-2\pi\eta_c}\left[a_c-(d-3)\eta_c^2-4\pi\eta_c a_1\right]
 \frac{\cos{[4\pi(r_c/\sqrt{f_c})\omega]}}{[4\pi(r_c/\sqrt{f_c})\omega]^2}
\right).
\end{eqnarray}
In (\ref{Struct_fine_asymp_dimD}), the first two terms correspond to
the eikonal description constructed in section 4 of
Ref.~\cite{DecaniniEspositoFareseFolacci2011} while the third,
fourth and fifth ones describe the fine structure of the absorption
cross section. Let us note that the arguments of the sine and cosine
functions involve the orbital period, $2\pi (r_c/\sqrt{f_c})=2\pi
b_c$, of a massless particle orbiting the black hole on the photon
sphere as well as its second harmonic. Of course, equation
(\ref{Struct_fine_asymp_dimD}) generalizes (\ref{Struct_fine_asymp})
for static spherically symmetric black holes and, {\it mutatis
mutandis}, the physical interpretation of the fine structure for the
Schwarzschild black hole provided in section 2 remains valid in the
general case. The coefficients $a_c$ and $a_1$ which appear in the
last two terms of (\ref{Struct_fine_asymp_dimD}) are defined in the
appendix (see equation (\ref{Residu_SSSBH_ac}) for $a_c$ and
equation (\ref{RPHF_an}) with $n=1$ for $a_1$) and are expressed in
terms of the derivatives of $f(r)$ taken on the photon sphere. It is
also important to note that the fifth term of
(\ref{Struct_fine_asymp_dimD}) was not present for the Schwarzschild
black hole. In fact, we have discarded it in section 2 because it
was numerically negligible for this particular black hole. As we
shall see later, it is also numerically insignificant for the five-
and six-dimensional Schwarzschild-Tangherlini black holes and for
the four-dimensional Reissner-Nordstr\"om black hole. However, we
consider that, in the general case, it cannot be discarded.

\bigskip

To conclude the general theory, let us remark that for ``very high
frequencies", we can eliminate the last two terms of
(\ref{Struct_fine_asymp_dimD}) and we have therefore at our disposal
a nice and simple formula describing accurately the absorption cross
section for a massless scalar field propagating on an arbitrary
static and spherically symmetric black hole.

\subsection{Application 1: Schwarzschild-Tangherlini black holes}

We now apply the general theory developed in the previous subsection
to Schwarzschild-Tangherlini black holes. They are generalization of
the four-dimensional Schwarzschild black hole constructed in the
sixties by Tangherlini \cite{Tangherlini63}. For a $d$-dimensional
Schwarzschild-Tangherlini black hole, we have
\begin{equation} \label{ST_fr}
f(r)=1-\left(\frac{r_h}{r}\right)^{d-3}.
\end{equation}
Here $r_h$, which denotes the standard coordinate of the event
horizon, is linked to the mass $M$ of the black hole by
\begin{equation}\label{r0dimd}
r_h^{d-3}=\frac{16\pi M}{(d-2){\cal A}_{d-2}}
\end{equation}
where ${\cal A}_{d-2}=2\pi^{(d-1)/2}/\Gamma[(d-1)/2]$ is the area
of the unit sphere $S^{d-2}$. The photon sphere is then located at
\numparts
\begin{equation}\label{rcSchTang}
r_c=r_h\left(\frac{d-1}{2}\right)^{1/(d-3)},
\end{equation}
the associated $\eta_c$ parameter is given by
\begin{equation}\label{etaSchTang}
\eta_c=\sqrt{d-3}
\end{equation}
while the corresponding critical impact parameter
$b_c=r_c/\sqrt{f_c}$ reads
\begin{equation}\label{bc_SchTang}
b_c= \sqrt{\frac{d-1}{d-3}}r_c.
\end{equation}
\endnumparts

\bigskip

For $d=5$, we have $\sigma_\mathrm{geo}=(4\pi/3)b_c^3$ with
$b_c=\sqrt{2}r_c$ and $r_c=\sqrt{2}r_h$ as well as $\eta_c=\sqrt{2}$
and the general formula (\ref{Struct_fine_asymp_dimD}) then reduces
to
\begin{eqnarray} \label{Struct_fine ST5}
\fl\sigma_\mathrm{abs}^\mathrm{ST \, d=5}(\omega)\approx
\sigma_\mathrm{geo}\left(1+12\sqrt{2}\pi e^{-\sqrt{2} \pi} \,
\frac{\sin [2\pi b_c \omega]}{2\pi b_c \omega}  + 24 \sqrt{2}\pi
e^{-2\sqrt{2} \pi} \, \frac{\sin [4\pi b_c \omega]}{4\pi b_c \omega}
\right. \nonumber \\
\fl \left. \qquad \qquad \qquad \qquad + 3\pi^2(13-\sqrt{2}\pi)
e^{-\sqrt{2} \pi} \, \frac{\cos [2\pi b_c \omega]}{(2\pi b_c
\omega)^2}
 \right) .
\end{eqnarray}
For $d=6$, we have $\sigma_\mathrm{geo}=(\pi^2/2)b_c^4$ with
$b_c=\sqrt{5/3}r_c$ and $r_c=(5/2)^{1/3}r_h$ as well as
$\eta_c=\sqrt{3}$ and the general formula
(\ref{Struct_fine_asymp_dimD}) then reduces to
\begin{eqnarray} \label{Struct_fine ST6}
\fl \sigma_\mathrm{abs}^\mathrm{ST \, d=6}(\omega)\approx
\sigma_\mathrm{geo}\left(1-16\sqrt{3}\pi e^{-\sqrt{3} \pi} \,
\frac{\sin [2\pi b_c \omega]}{2\pi b_c \omega} + 32\sqrt{3}\pi
e^{-2\sqrt{3} \pi} \, \frac{\sin [4\pi b_c \omega]}{4\pi b_c \omega}
\right. \nonumber \\
\fl \left. \qquad \qquad \qquad \qquad + 16\pi^2\frac{(-114+
5\sqrt{3}\pi)}{15} e^{-\sqrt{3} \pi} \, \frac{\cos [2\pi b_c
\omega]}{(2\pi b_c \omega)^2}
 \right).
\end{eqnarray}
It should be noted that, in (\ref{Struct_fine ST5}) and
(\ref{Struct_fine ST6}), we have discarded the fifth term of
(\ref{Struct_fine_asymp_dimD}) which we have found numerically
negligible.

\subsection{Application 2: The four-dimensional Reissner-Nordtr\"om black hole}

We also apply the general theory developed in subsection 3.1 to the
four-dimensional Reissner-Nordtr\"om black hole (see, e.g.,
Ref.~\cite{MTW}). In this case, we have
\begin{equation} \label{RN4_fr}
f(r)=1-\frac{2M}{r}+\frac{Q^2}{r^2}
\end{equation}
where $M$ is the mass of the black hole and $Q$ denotes its charge
and we shall assume that $M>Q$. This black hole has inner and outer
horizons located respectively at
\numparts
\begin{eqnarray}
&& r_{-}=M-\sqrt{M^{2}-Q^{2}},\\
&& r_{+}= M+\sqrt{M^{2}-Q^{2}} .
\end{eqnarray}
\endnumparts
We are interested only in the outer horizon with radius at
$r_h=r_{+}$ because we have $f(r)>0$ for $r \in ]r_h,+\infty[$. The
photon sphere is then located at \numparts
\begin{equation}\label{rc_RN4}
r_c=\frac{1}{2}(3M+\sqrt{9M^{2}-8Q^{2}}),
\end{equation}
(let us note that $r_c>r_h$) and the associated $\eta_c$ parameter
is given by
\begin{equation}\label{eta_RN4}
\eta_c=\sqrt{1-\frac{2Q^{2}}{r_c^{2}}}.
\end{equation}
The critical impact parameter $b_c=r_c/\sqrt{f_c}$ reads
\begin{equation}\label{bc_RN4}
b_c= \frac{\sqrt{3} r_c}{\sqrt{1- Q^{2}/r_c^{2}}}
\end{equation}
and we have for the corresponding geometrical cross section
\begin{equation}\label{SecGeom_RN4}
\sigma_\mathrm{geo}=\pi b_c^2= \frac{3\pi r_c^2}{1- Q^{2}/r_c^{2}}.
\end{equation}
\endnumparts

\bigskip

For this four-dimensional Reissner-Nordstr\"om black hole, the
general formula (\ref{Struct_fine_asymp_dimD}) then reduces to
\begin{eqnarray} \label{Struct_fine RN4}
\fl \sigma_\mathrm{abs}^\mathrm{RN \, d=4}(\omega)\approx
\sigma_\mathrm{geo}\left\{1-8\pi \eta_c e^{-\pi\eta_c}  \frac{\sin
[2\pi b_c \omega]}{2\pi b_c \omega}
+ 16 \pi \eta_c e^{-2\pi\eta_c}\frac{\sin [4\pi b_c \omega]}{4\pi b_c \omega}  \right. \nonumber \\
\fl \left. \qquad  -\frac{4\pi^2 e^{-\pi\eta_c}}{9\eta_c^4 }
\left[\left(13-\frac{72Q^2}{r_c^2}+\frac{123Q^4}{r_c^4}-\frac{82Q^6}{r_c^6}\right)
\nonumber \right.\right. \\
\fl \left. \left. \qquad \qquad \qquad \quad
-\frac{\pi\eta_c}{3}\left(7-\frac{18Q^2}{r_c^2}-\frac{39Q^4}{r_c^4}+\frac{50Q^6}{r_c^6}\right)
\right]\frac{\cos [2\pi b_c \omega]}{(2\pi b_c \omega)^2} \right\}.
\end{eqnarray}
In (\ref{Struct_fine RN4}), we have discarded the fifth term of
(\ref{Struct_fine_asymp_dimD}) which we have found numerically
negligible.

\section{Concluding remarks}

The existence of a simple fine structure in the ``absorption
spectrum" of black holes is an interesting feature which, to our
knowledge, has never been noted and which, furthermore, must be
certainly pointed out from a theoretical point of view. As was
already the case for the fluctuations around the capture cross
section of the black hole, this fine structure is only due to the
``surface wave" trapped near the photon sphere which is associated
with the first Regge pole. However, as emphasized at the end of
section 2, for the interpretation of the fine structure, we must now
take into account the multiple circumnavigations around the black
hole of this surface wave as well as its dispersive character. It is
moreover important to recall the duality existing between the Regge
poles and the complex frequencies of the weakly damped quasinormal
modes of the black hole (see
Refs.~\cite{DecaniniFJ_cam_bh,DecaniniFolacci2010a,
DecaniniFolacciRaffaelli2010b}). It could permit us to provide an
interpretation of the fine structure of the high-energy absorption
cross section for black holes in terms of quasinormal modes.

\bigskip

For the Schwarzschild black hole, the amplitude of the fine
structure is around $5 - 10 \, \%$ of that of the eikonal
fluctuations, so, in our opinion, its existence could even have
interesting physical consequences. This could be the case, for
example, in the context of strong gravitational lensing. Indeed, as
we have already noted in Ref.~\cite{DecaniniFolacci2010a}, until now
strong gravitational lensing has mainly been considered in the
framework of geometrical optics. A description based on wave
concepts and, in particular on Regge pole techniques, could furnish
a more correct description with new effects predicted. As we have
also remarked, the nonlinearities of the Regge trajectories could
induce possible observational consequences. Some
speculations/results contained in two recent papers
\cite{StefanovYazadjievGyulchev2010,WeiLiuGuo2011} seem also to
indicate that the resonant and absorption spectra of the
Schwarzschild black hole could play a crucial role in the context of
strong gravitational lensing and, in particular, that high-energy
absorption cross section and strong gravitational lensing are
intimately related. In that case, the eikonal and fine structures of
the absorption cross section of the Schwarzschild black hole could
be observed, perhaps even in the very near future, using the new
generation of experimental devices still under development (see,
e.g., Ref.~\cite{Eisenhauer_GRAVITY}) and designed in order to
explore the effect of spacetime curvature near the event horizon of
the supermassive black hole located at the Galactic Center.

\begin{figure}
\includegraphics[width=13.1cm]{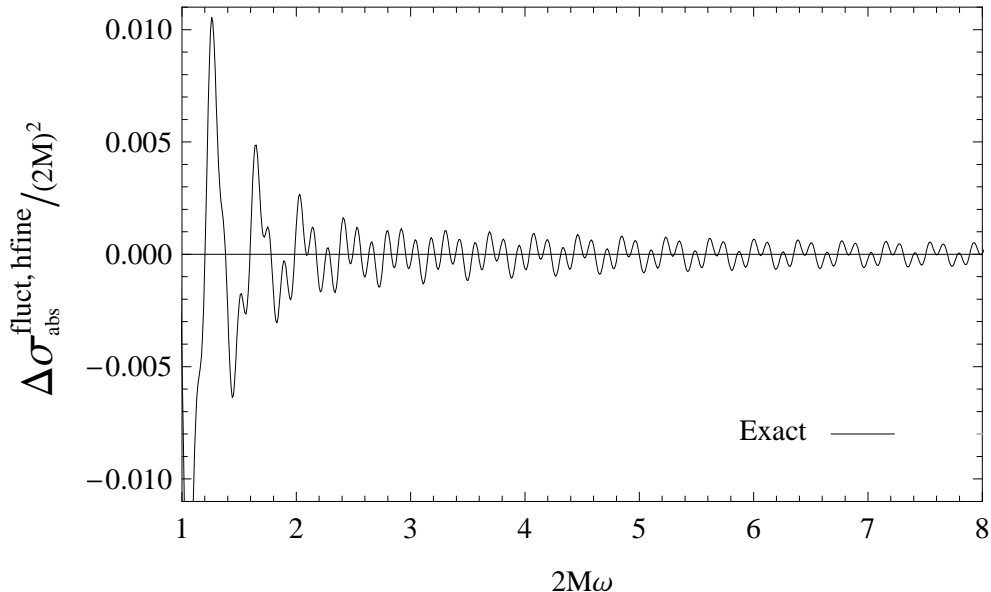}
\caption{\label{fig:HyperFineStructure}Hyperfine structure of the
total absorption cross section for a massless scalar field
propagating in the Schwarzschild geometry.}
\end{figure}

\bigskip

To conclude our paper, we invite the reader to look at figure
\ref{fig:HyperFineStructure} where we have displayed the ``hyperfine
structure" of the high-energy absorption cross section of the
Schwarzschild black hole. It has been obtained by subtracting to the
exact fine structure numerically constructed in section 2 the
contributions corresponding to the third and fourth terms of
(\ref{Struct_fine_asymp}) as well as the smooth contribution
$\pi/(12\sqrt{3}M\omega)^2$ discussed in section 2 of
Ref.~\cite{DecaniniEspositoFareseFolacci2011}. Even if the curve
displayed is a rather regular one (almost for high frequencies), it
presents a much more complicated behavior than the fine structure.
We think it could be described analytically by taking into account,
in addition to higher harmonics associated with the first Regge pole
[and in particular the  ($m=3$)-term of equation (\ref{devS sin_a})]
and to the nonlinearities associated with the first Regge pole and
its corresponding residue [higher order terms of equations
(\ref{PRapproxWKB_hf3}) and (\ref{RESapproxWKB_hf3})], the
contribution of the second Regge pole [the ($n=2$)-term of equation
(\ref{sigma_abs_SC3})]. However, because the amplitude of the
hyperfine structure is very weak in comparison with that of the fine
structure, it does not seem necessary (or even interesting from the
physical point of view) to achieve such a description.

\bigskip

\ack It is a pleasure to thank Gilles Esposito-Far\`ese for his
interest in our reggeization program of black hole physics and his
enthusiasm during the elaboration of our previous joint work on
absorption by black holes.

\appendix
\section*{Appendix}
\setcounter{section}{1}

The WKB approach developed a long time ago by Schutz, Will and Iyer
\cite{SchutzWill,Iyer1,Iyer2} (see also Ref.~\cite{WillGuinn1988}
for other related aspects) to determine the weakly damped
quasinormal frequencies of black holes proved very efficient in
order to construct high-frequency asymptotic expansions for the
black hole Regge poles (see our previous works
\cite{DecaniniFolacci2010a,DecaniniFolacciRaffaelli2010b}). Here, we
shall use and extend it to extract, from the black hole greybody
factors, high-frequency asymptotic expansions for the Regge pole
residues. We shall first consider the ordinary Schwarzschild black
hole and then generalize our study to arbitrary static spherically
symmetric black holes endowed with a photon sphere.

\subsection{Regge poles and residues of the greybody factors:
the Schwarzschild black hole.}

We first consider the case of the four-dimensional Schwarzschild
black hole. Even if, in section 2 of this paper, we are only
interested in the poles and residues of the greybody factors
associated with a scalar field theory, we shall here treat the more
general case of a field of spin $s$ with $s=0, 1 \, \mathrm{and} \,
2$ which satisfies the Regge-Wheeler equation, including therefore,
in addition to the scalar field theory ($s=0$), electromagnetism
($s=1$) and axial gravitational perturbations ($s=2$). From a
technical point of view, the general case does not present no more
difficulties than the scalar field case and our results could be
helpful in near future.

\bigskip

The wave equations for the scalar field, for the electromagnetic
field, and for the axial gravitational perturbations propagating on
the Schwarzschild black hole reduce, after separation of variables,
to the Regge-Wheeler equation (\ref{RW}) but now, instead of the
scalar Regge-Wheeler potential (\ref{EffectivePot_spin_0}), we must
consider the spin-dependent potential
\begin{equation}\label{EffectivePot_spin_s}
V_\ell(r) = \left(\frac{r-2M}{r} \right) \left[
\frac{(\ell+1/2)^2-1/4}{r^2} +\frac{2(1-s^2)M}{r^3}\right]
\end{equation}
and we must furthermore assume that the ordinary angular momentum
index  $\ell \in \mathbb{N}$ satisfies $\ell \ge s$. For $\omega>0$
and $\ell$ given, the partial radial amplitude $\phi_{\omega
\ell}(r_\ast)$ still satisfies the boundary conditions
(\ref{BC_Schw}).

\bigskip

Here, it is important to note that the Regge-Wheeler potential
$V_\ell(r_\ast)$ given by (\ref{EffectivePot_spin_s}) behaves as a
potential barrier and presents a maximum near the photon sphere of
the Schwarzschild black hole located at $r_c=3M$. Let us denote by
$r_0(\ell)$ the position of this maximum expressed in the radial
Schwarzschild coordinate and by ${(r_*)}_0(\ell)$ the corresponding
tortoise coordinate. From (\ref{EffectivePot_spin_s}) it is easy to
obtain
\begin{equation}\label{Max_Schwarzschild}
r_0(\ell)=3M\left[ 1 -\frac{(1-s^2)}{9(\ell+1/2)^2}
+ \underset{\ell+1/2 \to +\infty}{\cal O}\left(\frac{1}{(\ell+1/2)^4}\right)
\right]
\end{equation}
and to show that the peak of the Regge-Wheeler potential is given by
\begin{eqnarray}
V_0(\ell) &\equiv& \left. V_{\ell}(r_*)\right|_{r_*={(r_*)}_0(\ell)}
=\left. V_{\ell}(r)\right|_{r=r_0(\ell)} \label{hauteur_V_0a} \\
 &=&  \frac{(\ell+1/2)^2}{27M^2} +
\frac{-3+ 8(1-s^2)}{324 M^2}
+ \underset{\ell+1/2 \to +\infty}{\cal O}\left(\frac{1}{(\ell+1/2)^2}\right). \nonumber \\
& & \label{hauteur_V_0b}
\end{eqnarray}

\bigskip

For $\ell \in \mathbb{N}$ and $\omega
>0$ with $\omega^2 $ near the peak $V_0(\ell)$ of the Regge-Wheeler potential, we
can use, following Iyer, Will and Guinn
\cite{Iyer1,Iyer2,WillGuinn1988}, a third-order WKB approximation
for the greybody factors (\ref{Greybodyfactors}). We have
\begin{equation}\label{Greybodyfactors_1}
\Gamma_\ell(\omega)= \frac{1}{1+\exp[2\mathcal{S}_\ell(\omega)]}
\end{equation}
where
\begin{eqnarray}\label{Greybodyfactors_2}
\fl  \mathcal{S}_\ell(\omega)= \pi k^{1/2} \left[ \frac{1}{2} z_0^2
+ \left(\frac{15}{64} b_3^2-\frac{3}{16} b_4\right)z_0^4 \right.
\nonumber \\
\fl \qquad \qquad \qquad \left. +
 \left(\frac{1155}{2048} b_3^4-\frac{315}{256}b_3^2 b_4 +\frac{35}{64} b_3b_5
+\frac{35}{128} b_4^2-\frac{5}{32} b_6\right)z_0^6 \right] \nonumber \\
\fl \quad  + \pi k^{-1/2} \left[ \left(\frac{3}{16} b_4-\frac{7}{64}
b_3^2\right)-\left(\frac{1365}{2048} b_3^4-\frac{525}{256} b_3^2b_4
+\frac{95}{64} b_3b_5 +\frac{85}{128} b_4^2-\frac{25}{32} b_6\right)
z_0^2   \right]. \nonumber \\
\end{eqnarray}
Here we use the notations
\begin{eqnarray}\label{Greybodyfactors_3}
& & z_0 \equiv z_0(\ell,\omega) = \sqrt{2 \, \frac{\omega^2 -
V_0(\ell)}{V^{(2)}_0(\ell)}}, \\
& & k \equiv k(\ell) = -\frac{1}{2} V^{(2)}_0(\ell), \\
&& b_p \equiv b_p(\ell) = \frac{2}{p!}
\frac{V^{(p)}_0(\ell)}{V^{(2)}_0(\ell)} \quad \mathrm{for} \quad
p>2,
\end{eqnarray}
with $V_0(\ell)$ defined by (\ref{hauteur_V_0a}) and
\begin{equation} \label{V_0 et dersup}
V^{(p)}_0(\ell) \equiv \left. \frac{d^p}{{dr_*}^p }
V_{\ell}(r_*)\right|_{r_*={(r_*)}_0(\ell)} \quad \mathrm{for} \quad
p \ge 2.
\end{equation}

\bigskip

Even if the authors of Refs.~\cite{Iyer1,Iyer2,WillGuinn1988} have
obtained the previous formulas for $\ell \in \mathbb{N}$ and $\omega
>0$, they have moreover shown that their results are helpful in order to
obtain third-order WKB approximations for the weakly damped complex
quasinormal frequencies of black holes. As we have already remarked
in Refs.~\cite{DecaniniFolacci2010a} and
\cite{DecaniniFolacciRaffaelli2010b}, these same formulas are very
efficient in order to construct high-frequency asymptotic expansions
for the black hole Regge poles and, here, we shall use them to
extract, from the black hole greybody factors, high-frequency
asymptotic expansions for the Regge pole residues. First, we
consider that $\omega >0$ but we transform the angular momentum
$\ell$ appearing in the previous equations into a complex variable
$\lambda=\ell+1/2$ and we then consider the analytic extension
$\Gamma_{\lambda-1/2}(\omega)$ of the greybody factors
$\Gamma_\ell(\omega)$ defined by (\ref{Greybodyfactors}) and
(\ref{Greybodyfactors_1}) as well as the analytic extension
$\mathcal{S}_{\lambda-1/2}(\omega)$ of the ``phases"
$\mathcal{S}_\ell(\omega)$ given by (\ref{Greybodyfactors_2}). The
(Regge) poles of the greybody factors are the solutions
$\lambda_n(\omega)$ of the equation
\begin{equation}\label{RP_phase}
\mathcal{S}_{\lambda-1/2}(\omega)= i(n-1/2) \pi \quad \mathrm{with}
\quad n \in \mathbb{N}\setminus\lbrace{0\rbrace}
\end{equation}
(here we consider only those of the poles lying in the first
quadrant of the complex $\lambda$ plane) and it is easy to prove
that the corresponding residues are given by
\begin{equation}\label{Res_phase}
\gamma_n(\omega)= \frac{-1/2}{\left. [d \,
\mathcal{S}_{\lambda-1/2}(\omega) / d\lambda
]\right|_{\lambda=\lambda_n(\omega)}}.
\end{equation}
By inserting (\ref{Max_Schwarzschild}) into
(\ref{Greybodyfactors_2}) and considering the transformation $\ell
\rightarrow \lambda-1/2$, we obtain from (\ref{RP_phase}) the
asymptotic expansion
\begin{eqnarray}\label{PRapproxWKB_hf3}
\fl \lambda_n(\omega) = 3\sqrt{3} \, M\omega + i\left(n-1/2 \right)
+ \left(\frac{60 \left(n-1/2 \right)^2-144 (1-s^2) +115}{1296
\sqrt{3}}\right)\frac{1}{M\omega} \nonumber \\
\fl \quad - i \left(n-1/2 \right)\left(\frac{1220 \left(n-1/2
\right)^2-6912 (1-s^2) + 5555 }{419904}\right)\frac{1}{(M\omega)^2} \nonumber \\
\fl \quad + \underset{M\omega \to +\infty}{\cal O}\left(\frac{1}{(M\omega)^3}\right)
\end{eqnarray}
for the Regge poles (in agreement with the results of
Ref.~\cite{DecaniniFolacci2010a}). Finally, from
(\ref{Max_Schwarzschild}), (\ref{Greybodyfactors_2}),
(\ref{Res_phase}) and (\ref{PRapproxWKB_hf3}), we obtain for the
Regge pole residues
\begin{eqnarray}\label{RESapproxWKB_hf3}
\fl \gamma_n(\omega) = -\frac{1}{2\pi} + i
\left(\frac{5(n-1/2)}{108 \sqrt{3} \pi }\right)\frac{1}{M\omega}
\nonumber \\
\fl \quad +\left(\frac{3660 \left(n-1/2 \right)^2-6912 (1-s^2)
+5555}{839808 \pi }\right)\frac{1}{(M\omega)^2}
+ \underset{M\omega \to +\infty}{\cal O}\left(\frac{1}{(M\omega)^3}\right).
\nonumber \\
\fl
\end{eqnarray}
It should be noted that, in order to describe the fine structure of
the high-energy absorption cross section of the Schwarzschild black
hole, we need only the first three terms of (\ref{PRapproxWKB_hf3})
and the first two terms of (\ref{RESapproxWKB_hf3}). We have
provided one more higher order which could be helpful to describe
analytically the hyperfine structure of the high-energy absorption
cross section (see our final remark in section 4).

\subsection{Regge poles and residues of the greybody factors:
static spherically symmetric black holes}

We now focus our attention on the more general case of a static
spherically symmetric black hole of arbitrary dimension $d \ge 4$
endowed with a photon sphere which has been considered in section 3.
As noted in that section, a scalar field theory defined on such a
gravitational background, as the scalar field theory defined on the
Schwarzschild black hole, is governed by the Regge-Wheeler equation
(\ref{RW}) and the boundary conditions (\ref{BC_Schw}) but, now, the
Regge-Wheeler potential is no longer given by
(\ref{EffectivePot_spin_0}) but by the much more complicate
expression (\ref{RWPot_dimD}). However, due to the assumptions
listed at the beginning of section 3, the behaviors of
(\ref{EffectivePot_spin_0}) and (\ref{RWPot_dimD}) are quite
similar. As a consequence, {\it mutatis mutandis}, the calculations
of subsection A.1 of this appendix can be ``easily" generalized.

\bigskip

We first note that the Regge-Wheeler potential $V_\ell(r_\ast)$
given by (\ref{RWPot_dimD}) behaves as a potential barrier and
presents a maximum near the photon sphere of the black hole located
at $r=r_c$. Let us denote again by $r_0(\ell)$ the position of this
maximum expressed in the standard radial coordinate and by
${(r_*)}_0(\ell)$ the corresponding tortoise coordinate. From
(\ref{RWPot_dimD}) we can obtain
\begin{equation}\label{Max_SSSBH1}
\fl r_0(\ell)=r_c\left[ 1   +\frac{\delta_c}{[\ell+(d-3)/2]^2} +
\underset{\ell+(d-3)/2 \to +\infty}{\cal
O}\left(\frac{1}{[\ell+(d-3)/2]^4}\right) \right]
\end{equation}
with
\begin{equation}\label{Max_SSSBH2}
\delta_c=\frac{1}{2}(d-2) f_c \left(\frac{(d-2)f_c
+r_c^2f_c^{(2)}}{2f_c-r_c^2f_c^{(2)}}\right)
\end{equation}
which generalizes (\ref{Max_Schwarzschild}) and we can then show
that the peak of the Regge-Wheeler potential (\ref{RWPot_dimD})
which remains defined by (\ref{hauteur_V_0a}) is now given by
\begin{eqnarray}\label{DevRWPot_SSSBH}
&&V_{0}(\ell)=\frac{[\ell+(d-3)/2]^{2}}{r_{c}^{2}/f_c}
+\frac{\left[d(d-2)f_{c}-(d-3)^{2}\right]}{4r_{c}^{2}/f_{c}}
\nonumber\\
&&\qquad \qquad \qquad +\underset{\ell+(d-3)/2 \to +\infty}{\cal
O}\left( \frac{1}{[\ell+(d-3)/2]^2}\right)
\end{eqnarray}
which generalizes (\ref{hauteur_V_0b}). In equations
(\ref{Max_SSSBH2}) and (\ref{DevRWPot_SSSBH}) we have used the
notations (\ref{fc}) introduced at the beginning of section 3.

\bigskip

We can therefore consider that the WKB approximation
(\ref{Greybodyfactors_1})-(\ref{Greybodyfactors_2}) of subsection
A.1 remains valid in the more general case considered here. However,
in (\ref{Greybodyfactors_2}) we shall now take into account only the
terms of orders 1 and 2 of the WKB approximation, i.e., the terms in
$k^{1/2}z_0^2$, $k^{1/2}z_0^4$ and $k^{-1/2}$, in order to avoid
heavy calculations and because the description of the fine structure
of the absorption cross section can be fully achieved with that
precision. Then, by inserting (\ref{Max_SSSBH1}) into
(\ref{Greybodyfactors_2}) and using now the transformation $\ell
\rightarrow \lambda-(d-3)/2$, we obtain from (\ref{RP_phase}) the
asymptotic expansion
\begin{equation}\label{RPHF}
\fl \lambda_{n}(\omega)=\frac{r_{c}}{\sqrt{f_{c}}}~\omega
+i\eta_c(n-1/2)+\frac{a_{n}/2}{(r_{c}/\sqrt{f_{c}})\omega}
+\underset{(r_{c}/\sqrt{f_{c}})\omega \to +\infty}{\cal O}\left(
\frac{1}{[(r_{c}/\sqrt{f_{c}})\omega]^2}\right)
\end{equation}
\noindent with
\begin{eqnarray}\label{RPHF_an}
\fl a_{n}=-~\frac{1}{1152\eta_c^{4}}  \left\{
288f_{c}^{2}\left[(d^2-2d-1)f_{c}-(d-3)^{2}\right]\phantom{\left(f_{c}^{(2)}\right)^{3}}
\right. \nonumber \\
\fl \left.
 +144r_{c}^{2}f_{c}f_{c}^{(2)}\left[2(d-3)^{2}-(2d^2-4d-3)f_{c}\right]
- 72r_{c}^{3}f_{c}^{2}f_{c}^{(3)} \phantom{\left(f_{c}^{(2)}\right)^{3}} \right.\nonumber \\
\fl \left.  -18 r_{c}^{4}\left[4(d-3)^{2}
\left(f_{c}^{(2)}\right)^{2} -4(d-3)(d+1)
f_{c}\left(f_{c}^{(2)}\right)^{2} + f_{c}^{2}f_{c}^{(4)}\right]\right.\nonumber \\
\fl \left. + 36r_{c}^{5}f_{c}f_{c}^{(2)}f_{c}^{(3)} + \,
r_{c}^{6}\left[36\left(f_{c}^{(2)}\right)^{3}-7f_{c}\left(f_{c}^{(3)}\right)^{2}
+9f_{c}f_{c}^{(2)}f_{c}^{(4)}\right] \right\} \nonumber \\
\fl +(n-1/2)^{2}\frac{r_c^3f_c}{96\eta_c^{4}} \left\{24f_cf_c^{(3)}
+6r_cf_cf_c^{(4)}-12r_c^2f_c^{(2)}f_c^{(3)}
+r_c^3\left[5\left(f_c^{(3)}\right)^2-3f_c^{(2)}f_c^{(4)}\right]\right\} \nonumber \\
\fl
\end{eqnarray}
for the Regge poles (in agreement with our results in
Ref.~\cite{DecaniniFolacciRaffaelli2010b}). Here, in order to
simplify the results, we have used the notations (\ref{fc}) and
(\ref{Eta}) of section 3.

\bigskip

Finally, from (\ref{Max_SSSBH1}), (\ref{Greybodyfactors_2}),
(\ref{Res_phase}) and (\ref{RPHF}), we now obtain for the Regge pole
residues
\begin{equation}\label{Residu_SSSBH}
\fl
\gamma_n(\omega)=-\frac{\eta_c}{2\pi}+i\,\frac{(a_c/2\pi)(n-1/2)}{(r_c/\sqrt{f_c})\omega}
+\underset{(r_{c}/\sqrt{f_{c}})\omega \to +\infty}{\cal O}\left(
\frac{1}{[(r_{c}/\sqrt{f_{c}})\omega]^2}\right)
\end{equation}
\noindent with
\begin{eqnarray} \label{Residu_SSSBH_ac}
\fl a_c=\frac{r_c^3f_c}{96\eta_c^{4}}
\left\{24f_cf_c^{(3)}+6r_cf_cf_c^{(4)}
-12r_c^2f_c^{(2)}f_c^{(3)}+r_c^3\left[5\left(f_c^{(3)}\right)^2-3f_c^{(2)}f_c^{(4)}\right]\right\}.
\end{eqnarray}

\end{document}